# Derivations with Leibniz defect

**V. Kobelev**

Dept. Natural Sciences, University of Siegen, D-57076, Siegen, Germany email: kobelev@imr.mb.uni-siegen.de

**Abstract**

The non- Leibniz Hamiltonian and Lagrangian formalism is introduced in this article. The formalism is based on the generalized differentiation operator (κ-operator) with a non-zero Leibniz defect. The Leibniz defect of the introduced operator linearly depends on one scaling parameter. In a special case, if the Leibniz defect vanishes, the generalized differentiation operator reduces to the common differentiation operator. The κ-operator allows the formulation of the variational principles and corresponding Lagrange and Hamiltonian equations. The solutions of some generalized dynamical equations are provided closed form. With a positive Leibniz defect the amplitude of free vibration remains constant with time with the fading frequency ("red shift"). The negative Leibniz defect leads the opposite behavior, demonstrating the growing frequency ("blue shift"). However, the Hamiltonian remains constant in time in both cases. Thus the introduction of non-zero Leibniz defect leads to an alternative mathematical description of the conservative systems.

## Introduction

In the present article the alternative definition of the derivatives and the mechanical consequences of this alterative definition are investigated. The formalism is based on the generalized differentiation operator (κ-operator) with a non-zero Leibniz defect. The Leibniz defect of the generalized differentiation operator linearly depends on one scaling parameter. In a special case, if the scaling parameter turns to one, the Leibniz defect vanishes and generalized differentiation operator reduces to the common differentiation operator. The generalized differentiation operator allows the formulation of the variational principles and corresponding Lagrange and Hamiltonian equations.

## Generalized differentiation κ-operator with non-zero Leibniz defect

1°. A κ-differential field is a field $F$ with the generalized derivations

$$ð: F \to F$$

which satisfies the rules

$$ð(f+g) \to ð(f) + ð(g) \qquad (1)$$

and

$$ð(fg) \to ð(f)g + f\,ð(g) + \kappa\, f\, g. \qquad (2)$$

The last term is referred to as the Leibnitz defect [Tarasov, 2008, § 10.6]. The value $\kappa$ is a given constant of the inverse dimension of x. In the limit case $\kappa \to 0$ the Leibniz defect vanishes and generalized derivative turns into a common derivative $\partial$, which satisfies the Leibniz rule:

$$\partial(fg) \to \partial(f)g + f\partial(g).$$

2°. To define the generalized κ-derivation, we apply at first the power rule for the functions of the form $f = x^n$, whenever $n$ is a non-negative integer. The following κ-derivations are presumed:

$$ð1 = -\kappa, \qquad ðx = a \tag{3}$$

with a some constant $a$. Thus, the κ-derivations for non-negative integers we calculate by induction using the generalized Leibniz rule (2):

$$ð(x^n) = ð(x^{n-1})x + x^{n-1}ð(x) + \kappa x^{n-1}x = nax^{n-1} + (n-1)x^n\kappa. \tag{4}$$

The correspondence condition at $\kappa \to 0$ demands

$$ð(x^n)\Big|_{\kappa \to 0} = nax^{n-1} = \partial(x^n).$$

From the last identity the constant determines uniquely to $a = 1$. As a result, for all non-negative integers n the κ-derivations that satisfy (1),(2) and (3) are:

$$ð(x^n) = nx^{n-1} + (n-1)x^n\kappa. \tag{5}$$

The κ-differentiation is a linear operation on the space of differentiable functions. As the result, polynomials can also be differentiated using this rule. In this article the generalize derivations are applied to the derivations over the independent variables $x$ or $t$, for example:

$$ðf(t)/ðt = ð_t f \qquad \text{or} \qquad ðf(x)/ðx = ð_x f.$$

For brevity, in evident cases the subscript for the independent variable in the κ-derivative will be concealed.

3°. The next addressed question is the uniqueness of the derivative definition, based on product rule with the Leibniz defect. All solution operators of the classical Leibniz product rule were determined by [König, Milman, 2011]. The method established the cited article could be immediately extended to the rule (2) with the Leibniz defect. Similarly, the only solution of (2) that satisfies the correspondence condition (3) is essentially the κ-derivative. The product rule with the Leibniz defect after the substitution:

$$Tf = H(f(x))/H(x)$$

[König, Milman, 2011, Theorem 2] leads to the following functional equation:

$$\frac{F(fg)}{fg} = \frac{F(f)}{f} + \frac{F(g)}{g} + \kappa.$$

With the substitution

$$H(s) = F(\exp(s))$$

the product rule with the Leibniz defect reads:

$$H(s+t) = H(s) + H(t) + \kappa.$$

The only difference between the actual functional equation and the functional equation that is considered in [König, Milman, 2011] is the term $\kappa$. The solution of the actual functional equation delivers [Aczél, 1966]:

$$H(s) = s - \kappa.$$

The considerations of [König, Milman, 2011] are repeated for the product rule with the alternative entropy function that contains the non-vanishing Leibniz defect $\kappa$. Thus all solution operators of product rule with the Leibniz defect are determined.

## Adjoint generalized derivation operators

1°. In the first section we saw the algebraic definition of the non-Leibniz derivative and it computation based on the recurrences. We study now an alternative definition. The derivative of $f(x)$ with respect to $x$ is the function $\eth f$ and is defined as:

$$\eth f(x) = \lim_{\varepsilon \to 0} \frac{f(s^b x + \varepsilon)/s^a - f(x)}{\varepsilon}, \quad s = 1 + \varepsilon \cdot \kappa. \tag{6}$$

The two constants $a, b$ are yet arbitrary and have to be defined from the condition (2) and the correspondence principle $\eth x = 1$. The calculation (20) of the derivative for the function $x$ reduces to:

$$\eth x = \lim_{\varepsilon \to 0} \frac{(s^b x + \varepsilon)/s^a - x}{\varepsilon} = (b - a)\kappa x + 1,$$

such that the requirement $\eth x = 1$ demands $a = b$. Substitution of the expression (20) with two parameters $a, b$ in (2) leads to:

$$\eth(f)g + f\eth(g) + \kappa f g - \eth(fg) = (1-a)\kappa f(x)g(x). \tag{7}$$

Consequently, the condition (2) will be fulfilled, if $a = 1$. With this set of parameters the alternative definition of the $\kappa$-derivative reads:

$$\eth f(x) \stackrel{def}{=} \lim_{\varepsilon \to 0} \frac{f(sx + \varepsilon)/s - f(x)}{\varepsilon}, \quad s = 1 + \varepsilon \cdot \kappa.$$

This definition is equivalent to the algebraic definition (2) and establishes the relation between the common derivatives and their non-Leibniz pendants:

$$\eth f / \eth x \equiv (1 + \kappa x)f' - \kappa f, \quad f' = df/dx. \tag{8}$$

The relations (8) reduce the $\kappa$-differential equations to the ordinary differential equations.

4°. The adjoint $\ell^*$ of a given bounded operator $\ell$ acting on a certain Hilbert space K, with inner product, is defined by the following equality [Paycha, 2006]:

$$(\ell \varphi, \psi) = (\varphi, \ell^* \psi).$$

Here $\varphi$ and $\psi$ are arbitrary vectors in K. The role of inner product in mechanics play the integration:

$$\int \psi \cdot \ell \varphi \, dt = \int \varphi \cdot \ell^* \psi \, dt.$$

Namely, for the operator $\eth + \lambda$ the adjoint operator is $(\eth + \lambda) - 3\kappa\psi + 2\lambda\psi$, such that:

$$\int \psi \cdot (\eth + \lambda)\varphi \, dt = \int \psi \cdot (\eth + \lambda)\varphi \, dt = \int \varphi \cdot [(\eth + \lambda) - 3\kappa + 2\lambda]\psi \, dt . \tag{9}$$

If the term $(-3\kappa + 2\lambda)\psi$ vanishes, the resulting operator will be self-adjoint. The corresponding self-adjoint operator:

$$Đf \equiv (\eth + \tfrac{3}{2}\kappa)f \equiv (1 + \kappa x)f' + \tfrac{1}{2} f \tag{10}$$

with $\lambda = \tfrac{3}{2}\kappa$ will be applied for construction of Hamiltonian. The operator $\eth$, as follows from (9) with $\lambda = 0$, is not-self adjoint.

### Generalized κ-differential equations of the first and second order

1°. The special functions will be defined through the solution of κ-differential equations. At first, we define the κ-exponential function $\exp_\kappa(x, \omega)$ as the solution of the linear ordinary κ-differential equation of the first order:

$$\eth \exp_\kappa(x, \omega) = \omega \exp_\kappa(x, \omega). \tag{11}$$

The solution is searched in form of the Maclaurin series with so far unknown coefficients $e_n = e_n(\omega, \kappa)$:

$$\exp_\kappa(x, \omega) = \sum_{n=0}^\infty e_n x^n.$$

Using the rule (5), the κ-derivation of the Maclaurin series reduces to:

$$\eth \exp_\kappa(x, \omega) = \sum_{n=0}^\infty e_n \eth x^n = \sum_{n=0}^\infty e_n \left(nx^{n-1} + (n-1)x^n \kappa\right) \tag{12}$$

The recurrence for the coefficients $e_n$ appears after the substitution of the formulas (7) into (6) and collecting the equal powers of the variable $x$:

$$(n+1)e_{n+1} + (n-1)\kappa e_n = \omega e_n . \tag{13}$$

The common methods for solution of recurrences are demonstrated in [Everest et al 2003] and [Wimp, 1984]. The solution of the recurrence (8) reads:

$$e_n = e_0 \frac{(-\kappa)^n \Gamma(n - 1 - \omega/\kappa)}{\Gamma(-1 - \omega/\kappa)\Gamma(n+1)}, \tag{14}$$

The $e_0$ in (9) plays the role of the integration constant. The generating function $E(x)$ for the recurrence (9) satisfies the common ordinary differential equation:

$$(1+x\kappa)\frac{dE(x)}{dx} - \kappa E(x) = \omega E(x) \qquad \text{with } E(0)=1. \qquad (15)$$

From the condition $E(0)=1$ one gets $e_0 = 1$, such that the κ-exponential function appears in the form:

$$E(x) \equiv \exp_\kappa(x,\omega) = \sum_{n=0}^{\infty} \frac{(-\kappa x)^n \Gamma(n-1-\omega/\kappa)}{\Gamma(-1-\omega/\kappa)\Gamma(n+1)} \equiv (1+\kappa x)^{1+\omega/\kappa}. \qquad (16)$$

The graphs of the functions $\exp_\kappa(x,\omega)$ for different parameters of $\kappa$ and $\omega = 1$ are shown on the Fig.1.

The κ-exponential function reduces to the ordinary exponential function in its limit case equivalently, $\kappa \to 0$:

$$\lim_{\kappa \to 0}\exp_\kappa(x,\omega) = \exp(\omega x) \qquad \text{and} \qquad \exp_\kappa(x,-\kappa) = 1. \qquad (17)$$

The second identity demonstrates that for $\omega = -\kappa$ the $\exp_\kappa(x,\omega)$ turns into a constant 1. This unusual, but important property of the function $\exp_\kappa(x,\omega)$ is in accordance with the first equation in (3).

2°. At second, we define the κ-trigonometric functions. The definition is based on the Euler's formula:

$$\exp_\kappa(x,i\omega) = \cos_\kappa(x,\omega) + i\sin_\kappa(x,\omega). \qquad (18)$$

The corresponding κ-differential equations follow from Eq. (6) as its imaginary and real parts:

$$\eth\sin_\kappa(x,\omega) = \omega\cos_\kappa(x,\omega), \qquad \eth\cos_\kappa(x,\omega) = -\omega\sin_\kappa(x,\omega). \qquad (19)$$

From (11) and (13) follow the solutions of Eq. (14):

$$\cos_\kappa(x,\omega) = (1+\kappa x)\Xi(t,\omega), \qquad \sin_\kappa(x,\omega) = (1+\kappa x)\Psi(t,\omega). \qquad (20)$$

The new functions in the equations (20) are the following:

$$\Xi(t,\omega) = \cos(\kappa^{-1}\ln(1+\kappa t)\omega), \qquad \Psi(t,\omega) = \sin(\kappa^{-1}\ln(1+\kappa t)\omega), \qquad \Xi^2 + \Psi^2 = 1. \qquad (21)$$

If $\kappa$ runs from $0$ to $\infty$ and the solution displays an oscillating behavior with an increasing amplitude over $x$.

In contrast, if $\kappa$ runs from $-\infty$ to $0$, the solution is entirely different. In this case the solution demonstrate oscillatory tendency with the decreasing amplitude and increasing frequency. The frequency turns to infinity, as radius approaches the value $|\kappa^{-1}|$. Thus, the range of functions (16) and (20) is restricted to $x < -\kappa^{-1}$ if $\kappa$ is negative. It is simply no real-valued solution for the greater values $x$.

An alternative way to determine the κ-trigonometric functions is based on recurrence method. For this purpose, we assume the κ-trigonometric functions are the Maclaurin series with the so far unknown real coefficients $c_n = c_n(\omega,\kappa)$ and $s_n = s_n(\omega,\kappa)$:

$$\cos_\kappa(x,\omega) \equiv \sum_{n=0}^{\infty} c_n x^n, \qquad \sin_\kappa(x,\omega) \equiv \sum_{n=0}^{\infty} s_n x^n.$$

The recurrences for these coefficients that follow from (9) are:

$$c_n = \operatorname{Re} e_n(x,\omega), \qquad s_n = \operatorname{Im} e_n(x,\omega). \tag{22}$$

The κ- trigonometric functions reduce to the ordinary trigonometric function in its limit case:

$$\lim_{\kappa \to 0} \sin_\kappa(x,\omega) = \sin(\omega x), \qquad \lim_{\kappa \to 0} \cos_\kappa(x,\omega) = \cos(\omega x). \tag{23}$$

The graphs of the functions $\sin_\kappa(x,1)$ and $\cos_\kappa(x,1)$ for different parameters of $\kappa$ are shown on the Fig.2 and Fig.3.

**Higher order and fractional $\kappa$−derivatives**

With the expression (23), the $\kappa$−derivative of an integer order $n$ reads:

$$\frac{\eth^n f}{\eth x^n} \equiv \sum_{k=0}^{n} \binom{n}{k} (-\kappa)^k (1+\kappa x)^{n-k} \frac{d^{n-k} f}{dx^{n-k}}, \qquad \binom{\alpha}{\beta} = \frac{\Gamma(\alpha+1)}{\Gamma(\alpha-\beta+1)\Gamma(\beta+1)}. \tag{24}$$

Using the representation of $\kappa$−derivative in the form (24), the fractional analogue could be directly defined. Let $\alpha$ be the rational order of fractional derivative. The fractional generalization of (24) due to [Liouville, 1832, p. 117] turns into:

$$\frac{\eth^\alpha f}{\eth x^\alpha} \stackrel{def}{=} \sum_{k=0}^{\infty} \binom{\alpha}{k} (-\kappa)^k (1+\kappa x)^{\alpha-k} \frac{d^{\alpha-k} f}{dx^{\alpha-k}}. \tag{25}$$

The derivative in the right side in the expression (25) is the Riemann-Liouville of an order $\alpha - k$ [Samko et al 1993].

The fractional $\kappa$−derivative of the power functions could be immediately determined using (25). For this purpose recall the definition of the high order fractional derivative [Samko eta al 1993]. The fractional derivative of an order $\alpha$ of the function $f = x^\beta$ is:

$$\frac{d^\alpha x^\beta}{dx^\alpha} = \frac{\Gamma(\beta+1)}{\Gamma(\beta-\alpha+1)} x^{\beta-\alpha}.$$

The substitution of this expression into (25) leads to:

$$\frac{\eth^\alpha x^\beta}{\eth x^\alpha} = \sum_{k=0}^{\infty} \binom{\alpha}{k} (-\kappa)^k (1+\kappa x)^{\alpha-k} \frac{\Gamma(\beta+1) x^{\beta-\alpha+k}}{\Gamma(\beta-\alpha+k+1)} = (1+\kappa x)^\alpha \Gamma(\alpha+1) x^{\beta-\alpha} L_\alpha^{\beta-\alpha}\left(\frac{\kappa x}{1+\kappa x}\right). \tag{26}$$

In the Eq. (26) the symbol $L_\alpha^\beta(x)$ stays for a generalized Laguerre polynomial [Olver et al, 2010, Table 18.3.1]. Another equivalent expression of the fractional $\kappa$−derivative of the power functions in terms of hypergeometric function ${}_2F_1$ is:

$$\frac{\eth^\alpha x^\beta}{\eth x^\alpha} = \frac{\Gamma(\beta+1)}{\Gamma(\beta+1-\alpha)}(1+\kappa x)^\alpha x^{\beta-\alpha} {}_2F_1\left([-\alpha],[1+\beta-\alpha],\frac{\kappa x}{1+\kappa x}\right).$$

With these formulas the fractional $\kappa$—derivatives could be calculated for the functions that allow the Maclaurin series.

## Integration

Once a $\eth$ generalized derivation is defined, we look for a corresponding indefinite κ-integral. We define now the κ-indefinite integral (κ-antiderivative). The κ-indefinite integral represents a class of functions whose κ-derivative is the integrand:

$$\int_{(\kappa)} f(x)dx = F(x) \leftrightarrow f(x) \stackrel{def}{=} \eth F(\mathrm{x})/\eth \mathrm{x}. \qquad (27)$$

Consequently, the integration of the function $f(x)$ to get the antiderivative $F(x)$ is equivalent to the solution of the ordinary differential equation of the first order (27):

$$F(x) = \left(\int_0^x \frac{f(t)dt}{(\kappa t + 1)^2} + C\right)(\kappa t + 1), \qquad (28)$$

where $C = F(0)$ is an integration constant.

For example, we determine the integrals of the for the functions of the form $f = x^n$, whenever $n$ is a non-negative integer. Assuming the constant $C$ to be zero, the integral of the function $x^n$ expresses as:

$$F_n(x) \stackrel{def}{=} \left(\int_0^x \frac{t^n dt}{(\kappa t + 1)^2}\right)(\kappa t + 1) = \frac{n(\kappa x + 1)x^n}{2\kappa}\Phi_{1,\frac{n}{2}}(x^2\kappa^2) - \frac{n(\kappa x + 1)x^{n+1}}{2}\Phi_{1,\frac{n+1}{2}}(x^2\kappa^2) - \frac{x^n}{\kappa}.$$

Here $\Phi_{a,\nu}(z)$ is the Lerch Phi-function [Olver et al, 2010]:

$$\Phi_{a,\nu}(z) = \sum_{m=0}^\infty \frac{z^m}{(\nu+m)^a}.$$

The Taylor series of the function $F_n$ reads:

$$F_n(x) = \sum_{m=0}^\infty \frac{(-1)^m n x^{1+m+n} \kappa^m}{(m+n)(m+n+1)}.$$

If $\kappa = 0$ all terms in the Taylor series, except the first one, vanish. Consequently, for all positive integers $n$ the κ-indefinite integral turns into the common indefinite integral:

$$\lim_{\kappa \to 0} F_n(x) = \lim_{\kappa \to 0} \int_{(\kappa)} x^n dx = \int x^n dx \equiv \frac{x^{n+1}}{n+1}.$$

For example, for $n=1$ we have:

$$\Phi_{1,\frac{1}{2}}(x^2\kappa^2) = \frac{2\operatorname{arctanh}(\kappa x)}{\kappa x} \xrightarrow[\kappa\to 0]{} 2, \qquad \Phi_{1,1}(x^2\kappa^2) = -\frac{\ln(1-\kappa^2 x^2)}{\kappa^2 x^2} \xrightarrow[\kappa\to 0]{} 1,$$

$$F_1(x) = \kappa^{-2}(x\kappa+1)\ln(1-\kappa^2 x^2)/2 + \kappa^{-2}(x\kappa+1)\operatorname{arctanh}(\kappa x) - x/\kappa \xrightarrow[\kappa\to 0]{} x^2/2 \ .$$

Since the κ-integration is a linear operation on the space of integrable functions, polynomials can also be integrated using the rule (27).

## Generalized Lagrangian and Hamiltonian

1°. In mechanics the time $t$ plays frequently the role of the independent variable $x$ and marks the evolution of the system. There are typically a number of dependent generalized coordinates $q_i, i=1,...,N$ instead of $f(x)$. A common mechanical system possesses the kinetic energy $T_0 = T_0(t, q_i, \dot{q}_i)$ and the potential energy $V = V(q_i)$. The common velocity $\dot{q}_i$ is the ordinary time derivative of the coordinate $q_i$. The Lagrangian of the common mechanical system is:

$$L_0 = L_0(t, q_i, \dot{q}_i) = T_0(t, q_i, \dot{q}_i) - V(q_i).$$

The lover Latin indices are used for generalized coordinates. . The correspondence principle is the following. The ordinary derivatives $\dot{q}_i$ must be replaced by $\dot{q}_i \to Ðq_i$ in the expression of the Lagrangian:

$$L = L(t, q_i, Ðq_i) = L(t, q_i, (1+\kappa t)\dot{q}_i + \tfrac{1}{2} q_i) = T(t, q_i, Ðq_i) - V(q_i)(q_i). \tag{29}$$

This function will be referred to as the κ-Lagrange function of the system. Application of the common Euler equations

$$d(\partial L/\partial \dot{q}_i)/dt - \partial L/\partial q_i = 0, \qquad i=1,...,N$$

to the Lagrangian $L(t, q_i, (1+\kappa t)\dot{q}_i + \tfrac{1}{2} q_i)$ (29) delivers $n$ coupled Euler-Lagrange equations:

$$Ð\{\partial L/\partial(Ðq_i)\} - \partial L/\partial q_i = 0, \qquad i=1,...,N. \tag{30}$$

The basic method in the Lagrange mechanics consists in treating the generalized coordinates as independent variables. The time dependence of these variables is determined by the Lagrange equations of the second order. The generalized velocities $Ðq_i$ are all dependent, derived quantities. The initial values for $q_i$ and $Ðq_i$ are determined from the $2n$ integration constants.

2°. The canonical momenta $p^i(t)$ are used instead of $Ðq_i$ which are defined by the ordinary partial (functional) derivatives of Lagrangian:

$$p^i \stackrel{def}{=} \partial L/\partial(Ðq_i), \qquad i=1,...,N. \tag{31}$$

For canonical momenta the upper Latin indices will be used. Then the momenta $p^i$ are raised at the same level with the coordinates $q_i(t)$. Consequently the set

$$\{q_i(t),\ p^i(t), i=1,\ldots,n\}$$

forms a sequence of $2n$ independent variables. These variables fulfil a sequence of $2n$ coupled differential equations of the first order:

$$Ðq_i = \partial H/\partial p^i, \qquad Ðp^i = -\partial H/\partial q_i, \qquad H = \sum_{i=1}^{n} p^i Ðq_i - L. \tag{32}$$

In the following we imply the convention for a summation about repeated indexes. The variables $q$ or $p$ without an index designate the whole sequence.

## Linear harmonic oscillator

Consider the system performing linear oscillations. For example, consider a system with one degree of freedom in a stable equilibrium position. Let $q_0$ be the value of the generalized coordinate corresponding to the equilibrium position. When the system is slightly displaced to a position $q$ from the equilibrium position, a force occurs which acts to restore the equilibrium when its potential energy. The potential energy can be written as

$$V(q) = Kq^2/2. \tag{33}$$

The coefficient $K$ represents the value of the second derivative of $V(q)$ for $q = 0$. The $\kappa$–kinetic energy of a system is assumed to be

$$T = M\,ðq^2/2. \tag{34}$$

Let the mass $M$ of the particle is one. The harmonic oscillator with one degree of freedom possesses the Lagrangians:

$$L_0 = L_0(t,q,\dot q) = T_0(t,q,\dot q) - V(q) = \dot q^2/2 - \omega^2 q^2/2, \qquad \omega^2 = K/M, \tag{35}$$

$$L = L(t,q,Ðq) = T(t,q,Ðq) - V(q) = Ðq^2/2 - \omega^2 q^2/2,\ H = (p^2 + \omega^2 q^2)/2. \tag{36}$$

The equations of motion of the $\kappa-$ harmonic oscillator reduce with (32) to:

$$Ðq = p \qquad Ðp = -\omega^2 q. \tag{37}$$

From the Eq. (37) follow the equation of motion:

$$Ð^2 q + \omega^2 q = 0 \qquad \text{or} \qquad ð^2 q + 3\kappa\,ðq + (\omega^2 + \tfrac{9}{4}\kappa^2)q = 0. \tag{38}$$

The second Eq. (38) looks on the first glance as the equation for the damped harmonic oscillator [Landau, Lifshitz, 1976]:

$$\partial^2 q + 2\lambda \partial q + \omega^2 q = 0. \tag{39}$$

The factor $3\kappa$ in Eq. (38) stays on place the damping coefficient $2\lambda$ in Eq. (39). The substitution of (23) into (54) leads to the ordinary differential equations:

$$(1+\kappa t)^2 q'' + 2\kappa(\kappa t + 1)q' + \Omega^2 q = 0, \qquad q(0) = q_0, q'(0) = q_1. \tag{40}$$

Here is $\Omega^2 = \omega^2 + \tfrac{1}{4}\kappa^2$. The solution of the differential equation (40) leads to the following expressions:

$$q = \frac{2\omega\Xi + \kappa\Psi}{2\omega\sqrt{1+\kappa t}} q_0 + \frac{\Psi}{\omega\sqrt{1+\kappa t}} q_1, \qquad p = \frac{\kappa\Xi - 2\omega\Psi}{2\sqrt{1+\kappa t}} q_0 + \frac{\Xi}{\sqrt{1+\kappa t}} q_1. \tag{41}$$

The functions $\Xi(t,\omega), \Psi(t,\omega)$ were defined in Eq. (16). If $\kappa < 0$ the wave length slopes down with time and vanishes at the moment of time $t = -\kappa^{-1}$. The point $t = -\kappa^{-1}$ is the reflection point. The period fades also as the logarithmic function of time ("blue shift"), but the amplitude remains constant in time. Moreover, the alternation of the wave length does not depend upon the wave length. Consequently, there is no dispersion of waves with the different initial lengths. The case $\kappa > 0$ leads to the increasing wave length with time. In this case the period is swells with time such that the "red shift" happens. The amplitude does not alter in this case as well.

In the limit case $\kappa \to 0$ the functions reduce to the standard trigonometric functions:

$$\Psi(t,\Omega) \to \sin(\omega t), \qquad \Xi(t,\Omega) \to \cos(\omega t).$$

**Poisson bracket and conservation laws in the non-Leibniz mechanics**

In general, the Poisson bracket of any two dynamical variables $f(t,q_i,p^i)$ and $g(t,q_i,p^i)$ is defined as

$$[f,g] = \sum_{i=1}^{n} \left( \frac{\partial f}{\partial q_i}\frac{\partial g}{\partial p^i} - \frac{\partial f}{\partial p^i}\frac{\partial g}{\partial q_i} \right). \tag{42}$$

For arbitrary functions $f(t,q_i,p^i)$, $g(t,q_i,p^i)$ and $h(t,q_i,p^i)$ the following relations are valid:

$$[f,g] = -[g,f], \tag{43}$$

$$[af,g] = a[f,g] \text{ with an arbitrary constant } a, \tag{44}$$

$$[f+h,g] = [f,g] + [h,g], \tag{45}$$

$$[[f,g],h] + [[g,h],f] + [[h,f],g] = 0, \tag{46}$$

$$\frac{\eth}{\eth t}[f,g] = \left[\frac{\eth f}{\eth t},g\right] + \left[f,\frac{\eth g}{\eth t}\right] + \kappa[f,g]. \tag{47}$$

The relations (43..46) do not involve the κ-derivatives and are well-known from mechanics [Sussman, Wisdom, 2014, §3.2] or [Marsden, Ratiu, 2010]. The relation (47) contains an additional term. The specific for the non-Leibniz mechanics identity (47) could be immediately demonstrated with the definition of the κ-derivative (2). From Hamilton equations (44) the κ-derivative of a function $F(t, q_i, p^i)$ becomes:

$$Đ F = \frac{\partial F}{\partial t} + \sum_{i=1}^{n} \frac{\partial F}{\partial q_i} Đ q_i + \frac{\partial F}{\partial p^i} Đ p^i = \frac{\partial F}{\partial t} + \sum_{i=1}^{n} \frac{\partial F}{\partial q_i}\left(\frac{\partial H}{\partial p^i}\right)\frac{\partial H}{\partial p^i} + \frac{\partial F}{\partial p^i}\left(-\frac{\partial H}{\partial q_i}\right) = [F, H] + \frac{\partial F}{\partial t}. \quad (48)$$

A special case of (48) is:

$$Đ q_i = [q_i, H], \qquad Đ p^i = [p^i, H] \quad (49)$$

The equations (49) are referred to as equations of motion (32) in Poisson bracket form.
Another special case is:

$$Đ H = [H, H] + \frac{\partial H}{\partial t} = \frac{\partial H}{\partial t}. \quad (50)$$

The identity $[H, H] = 0$ is used in (50). If the Hamiltonian does not depend directly on time $\partial H / \partial t = 0$, the subsequent formula demonstrates the time dependence of the Hamiltonian:

$$Đ H = 0. \quad (51)$$

This condition is the non-Leibnizian counterpart to the common conservation law $\partial H = 0$. In accordance with (10) we have $Đ\{(1+\kappa t)^{-1/2}\} = 0$, such that the solution of (51) reads $H(t) = H(t=0)/\sqrt{1+\kappa t}$.

## Relativistic non-Leibniz equations

The difference between the non-relativistic and the relativistic formalism is the Lorentz invariant notation. The four coordinates

$$(x_0, x_1, x_2, x_3) = (ct, -x, -y, -z)$$

form the components of a first order four-dimensional Cartesian tensor of 4-vector [Landau, Lifshitz, 1972, §38]. The Greek indices run from 0 to 3: $\mu = 0,1,2,3$. The Latin indices run form 1 to 3. In this notation $c$ is the speed of the speed of light. The time and space variables will be considered hereafter as normalized such that $c = 1$ and consequently [Landau, Lifshitz, 1972, Eq. (38.5)]:

$$g_{\mu\nu} = g^{\mu\varpi} = diag\ (1, -1, -1, -1).$$

The Lorentz transformation represents an orthogonal, length preserving transformation of the components of the 4-vector. An interval $d\tau$ symbolizes the length of the 4-vector [Landau, Lifshitz, 1972, Eq. (34.4)] uses the letter "s" for the interval]. The 4-vector components of $\kappa$-speed are:

$$v_\mu \stackrel{def}{=} Đ x_\mu,$$

In relativistic formulation the operator ð symbolizes the interval derivative with respect to $\tau$. The covariant and contravariant components of 4-vector $\kappa_\mu$ are assumed to be:

Version 05-12-2017            11

$$\kappa_\mu = (\kappa, -\kappa, -\kappa, -\kappa), \qquad \kappa^\mu = (\kappa, \kappa, \kappa, \kappa).$$

This assumption is logically motivated by the Lorentz invariance of the $\kappa-$differentiation and the previous normalization for speed of light. The quantities:

$$p^\mu \overset{def}{=} -\partial(\gamma L)/\partial(\text{Ð}x_\mu). \qquad\qquad \gamma = \sqrt{1 - v_1^2 - v_2^2 - v_3^2},$$

are regarded as the components of 4- dimensional momentum $\kappa$-vector similar to [Landau, Lifshitz, 1972, Eq. (39.2)]. The Lagrange equations from (38) convert to the covariant form:

$$\frac{\partial \gamma L}{\partial x_\mu} - \text{Ð}\frac{\partial \gamma L}{\partial v_\mu} = 0.$$

For example, for a free particle of a mass $M$ we have $\gamma L = -M\sqrt{v_\mu v^\mu}$. The self-adjoint operators

$$\text{Ð}_\mu = \eth/\eth x^\mu + \tfrac{3}{2}\kappa_\mu, \qquad\qquad \text{Ð}^\mu = \eth/\eth x_\mu + \tfrac{3}{2}\kappa^\mu$$

transform as the covariant and contravariant components of the 4–vector respectively. The equations of field $A^\nu$ could be derived from the assumption of a Hamilton's principle:

$$\delta J[A^\nu] = \delta \int \Im(x^\mu, A^\nu, \text{Ð}_\mu A^\nu) d^4 x^\mu, \qquad\qquad \int \Im d^3 x^k = L. \tag{52}$$

The function $\Im$ is the $\kappa-$Lagrangian field density. The common procedure uses of the modified field tensor of [Landau, Lifshitz, 1972, Eq. (50.1)]:

$$F^{\mu\nu} = \text{Ð}^\mu A^\nu - \text{Ð}^\nu A^\mu. \tag{53}$$

The expression of $\kappa-$Lagrangian density in terms of the modified field tensor is similar to the common form [Landau, Lifshitz, 1972, Eq. (53.4)]:

$$\Im = -\tfrac{1}{4} F^{\mu\nu} F_{\mu\nu}. \tag{54}$$

The application of the equations of the previous section to quadratic function (54) leads to the non-Leibniz variant of the wave equation in vacuum:

$$\Diamond A^\nu = 0, \qquad \Diamond = \text{Ð}_t^2 - \text{Ð}_x^2 - \text{Ð}_y^2 - \text{Ð}_z^2, \tag{55}$$

with the symbol $\Diamond$ for the non-Leibniz D'Alembert operator. The non-Leibnizian form of the Maxwell equations in vacuum are:

$$\text{Ð}_\lambda F_{\mu\nu} + \text{Ð}_\mu F_{\nu\lambda} + \text{Ð}_\nu F_{\lambda\mu} = 0, \qquad\qquad \text{Ð}_\mu F^{\mu\nu} = 0.$$

# Wave equation

1°. The general solution of the homogeneous wave equation $\Diamond\phi = 0$ (55) in the case $N=1$ can be found by means of characteristics method [Nettel, 2009]:

$$\phi(x,t) = \left(\kappa^{-1} + x\right)^{2(1-\lambda/\kappa)} F_1\left(\left(\kappa^{-1} + t\right)/\left(\kappa^{-1} + x\right)\right) + F_2\left(\left(\kappa^{-1} + t\right)\cdot\left(\kappa^{-1} + x\right)\right), \quad (56)$$

The initial conditions $\phi|_{t=0} = \phi_1(x)$, $\partial\phi/\partial t|_{t=0} = \phi_2(x)$ determine the functions $F_1, F_2$.

2°. The radial Green functions of the homogeneous wave equation will be found in spaces of dimensions $N = 1,2,3$ [Ohtaka, 2005]. Separation of variables is applied by making a substitution of the form $\phi(t,r) = \tau(t)\rho(r)$, breaking the resulting equation into two independent ordinary differential equations for radius- and time-depending functions:

$$\rho'' = -\frac{(N+1)\kappa r + N - 1}{(1+\kappa r)r}\rho' - \frac{2\Omega^2 r + (N-1)\kappa}{2(1+\kappa r)^2 r}\rho, \quad (57)$$

$$\ddot{\tau} = -\frac{2\kappa}{1+\kappa t}\dot{\tau} + \frac{(N-1)\kappa^2 - 2\Omega^2}{(1+\kappa t)^2}\tau. \quad (58)$$

The frequency $\Omega$ plays the role of the separation constant. The Eq. (57) and (58) the homogeneous hypergeometric differential equations [Morse, Feshbach, 1953]. The solution of the differential equation (57) leads to the expression for $\rho(r)$ as the function of radius $r$ to the singular source of waves:

$$\begin{aligned}
\rho(r) &= C_{11}\rho_1(r)r^{2-N} + C_{21}\rho_2(r) \\
\rho_1 &= (1+\kappa r)^{-(\kappa+Q)/2\kappa} {}_2F_1([z_1, z_2][3-N], -\kappa r), \\
\rho_2 &= (1+\kappa r)^{-(\kappa+Q)/2\kappa} {}_2F_1([z_3, z_4][N-1], -\kappa r), \\
Q &= \sqrt{(2N-1)\kappa^2 - 4\Omega^2}, \quad P = \sqrt{N^2\kappa^2 - 4\Omega^2} \\
z_1 &= -\frac{Q+(N-3)\kappa - P}{2\kappa}, z_2 = -\frac{Q+(N-3)\kappa + P}{2\kappa}, \\
z_3 &= \frac{-Q+(N-1)\kappa - P}{2\kappa}, z_4 = \frac{-Q+(N-1)\kappa + P}{2\kappa}.
\end{aligned} \quad (59)$$

Here ${}_2F_1$ are the hypergeometric functions [Olver, 2010]. In the limit case $\kappa \to 0$ the solution (59) reduces to an ordinary solution for the two-dimensional waves in terms of Bessel functions:

$$\rho(r) = C_{11}r^{1-N/2}J_{N/2-1}(\Omega r) + C_{21}r^{1-N/2}Y_{N/2-1}(\Omega r). \quad (60)$$

3°. For $N=1$ the general solution (59) simplifies to the elementary functions:

$$\rho(r) = C_{11}(r+\kappa^{-1})^{\left(-\kappa+\sqrt{\kappa^2-4\Omega^2}\right)/2\kappa} + C_{21}(r+\kappa^{-1})^{\left(-\kappa-\sqrt{\kappa^2-4\Omega^2}\right)/2\kappa}. \quad (61)$$

As $\kappa \to 0$ and $N = 1$ the solution (60) reduces to a solution of one-dimensional waves [Morse, Ingard, 1968]:

$$\rho(r) = C_{11}\sin(\Omega t) + C_{21}\cos(\Omega t). \quad (62)$$

As $\kappa \to 0$ and $N = 2$ the solution (60) reduces to an ordinary solution for the two-dimensional waves [Morse, Ingard, 1968]:

$$\rho(r) = C_{11} J_0(\Omega r) + C_{21} Y_0(\Omega r). \tag{63}$$

The spherical waves appear from (59) after the substitution of $N = 3$. As $\kappa \to 0$ the solution (59) reduces to a common solution for spherical waves [Morse, Ingard, 1968]:

$$\rho(r) = C_{11} r^{-1} \sin(\Omega r) + C_{21} r^{-1} \cos(\Omega r).$$

The solution of the second, time-depending equation (58) in the following form:

$$\tau(t) = C_{12}(t + \kappa^{-1})^{\left(-\kappa + \sqrt{(2N-1)\kappa^2 - 4\Omega^2}\right)/2\kappa} + C_{22}(t + \kappa^{-1})^{\left(-\kappa - \sqrt{(2N-1)\kappa^2 - 4\Omega^2}\right)/2\kappa}, \tag{64}$$

with the corresponding ordinary limit case:

$$\tau(t) = C_{12} \sin(\Omega t) + C_{22} \cos(\Omega t) \qquad \text{as} \qquad \kappa \to 0.$$

The plots of wave functions $\tau(t)$ for dimension N = 1 are shown on Figs.4. The wave functions $\tau(t)$ for a positive $\kappa$ are drawn for different initial phases. For the waves with a positive $\kappa$ the red shift occurs and the wave length increases with the distance from the source. The periods of waves also increase with the positive $\kappa$ with the distance from the source, as the wave propagates with the constant speed. The wave functions $\tau(t)$ for a negative $\kappa$ are drawn on Fig. 5. For the waves with a negative $\kappa$ the blue shift occurs and the wave length fades with the distance from the source. The maximal distance of wave propagation is the absolute value of $\kappa$, such that the full reflection of a wave takes place from the surface with the radius $|\kappa|$. For both cases the non-Leibnizian form of conservation law (51) takes place.

## Conclusions

The method for study of dynamical systems is based on the introduction of a derivative with the Leibniz defect. The differential algebra of this generalized derivative is briefly established. The simple non-Leibniz differential equations and their solutions are solved in closed form in terms of special functions. The introduction of generalized non-Leibniz derivative allows the formulating of the Newtonian and relativistic dynamical equations with Lagrange and Hamilton structure. The analytical solution of the equations of non-Leibniz oscillator is found. As an example for a solution of the partial differential equation with the generalized derivatives the $\kappa$−wave-equation is studied.

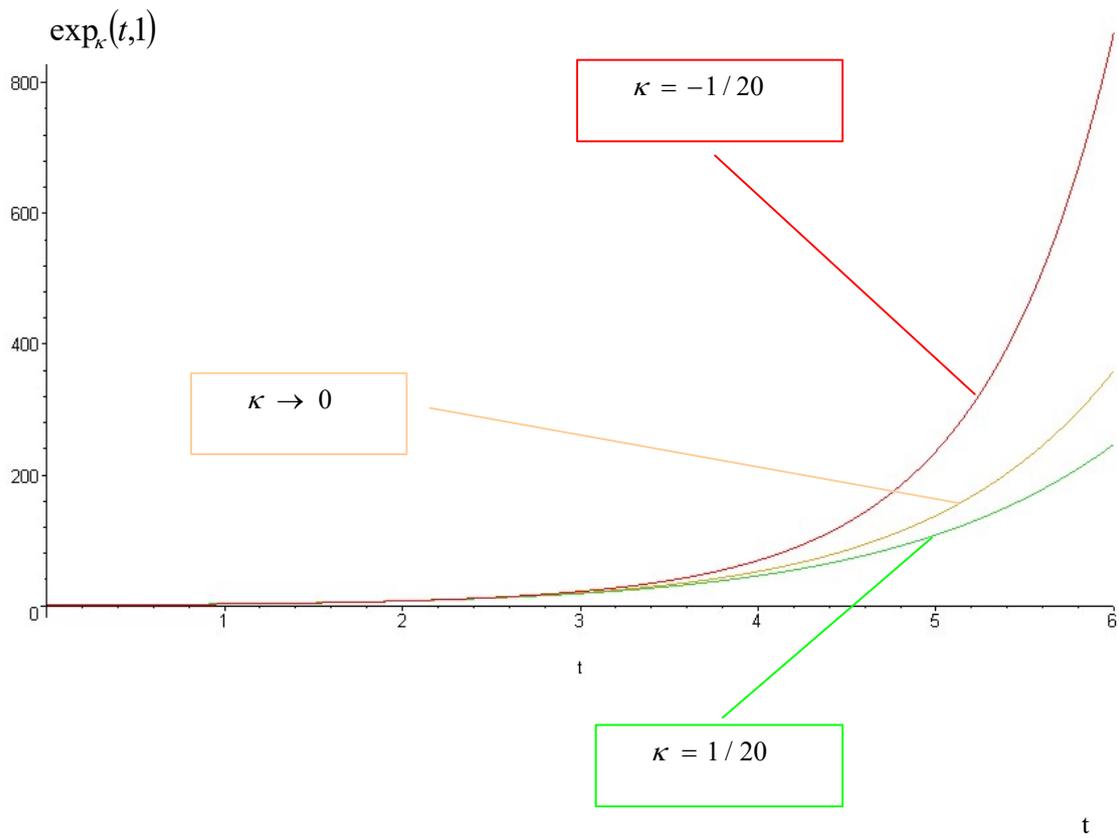

Fig 1. Functions $\exp_\kappa(t,1)$ different values of $\kappa$

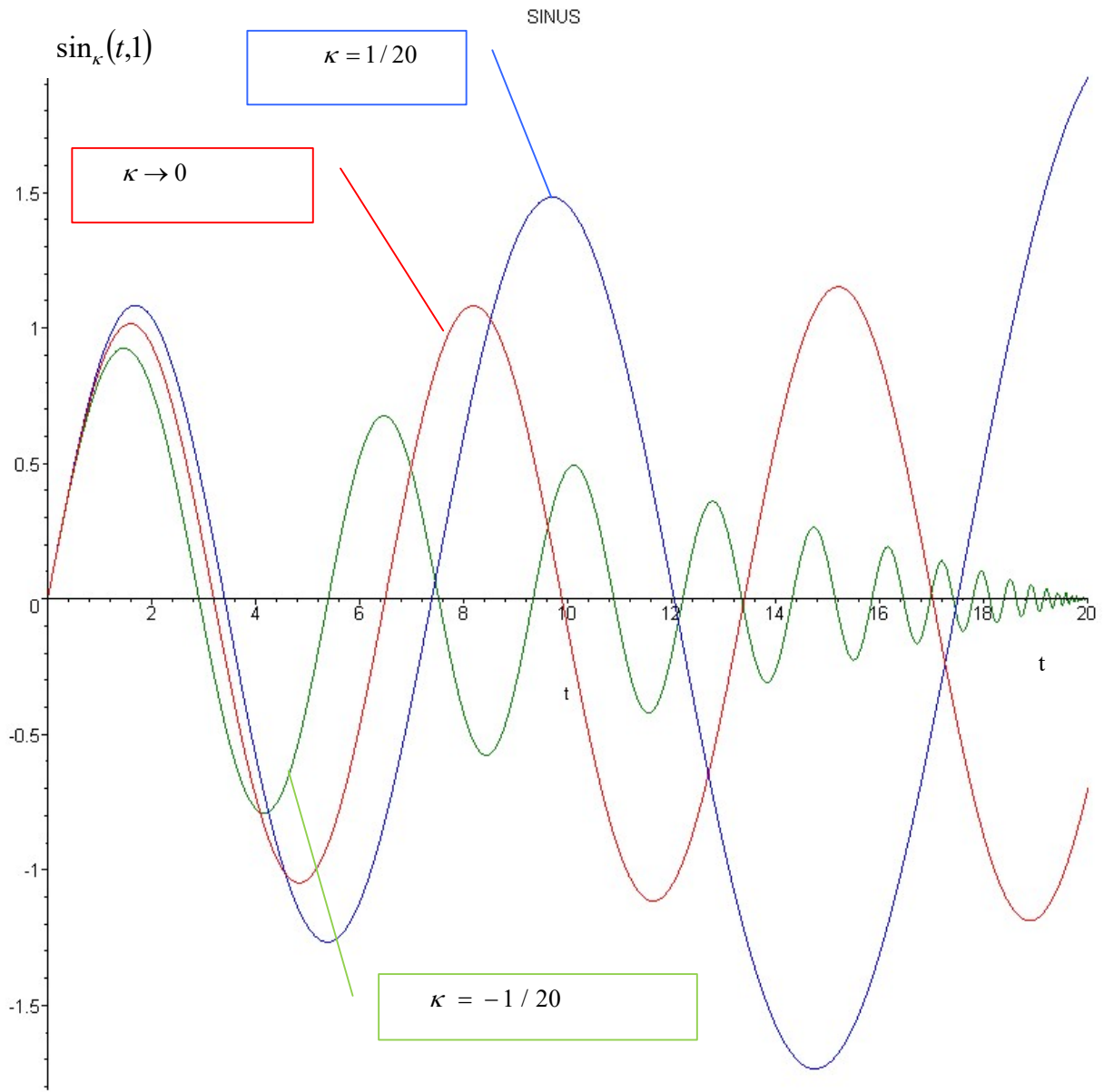

Fig 2. Functions $\sin_\kappa(t,1)$ different values of $\kappa$

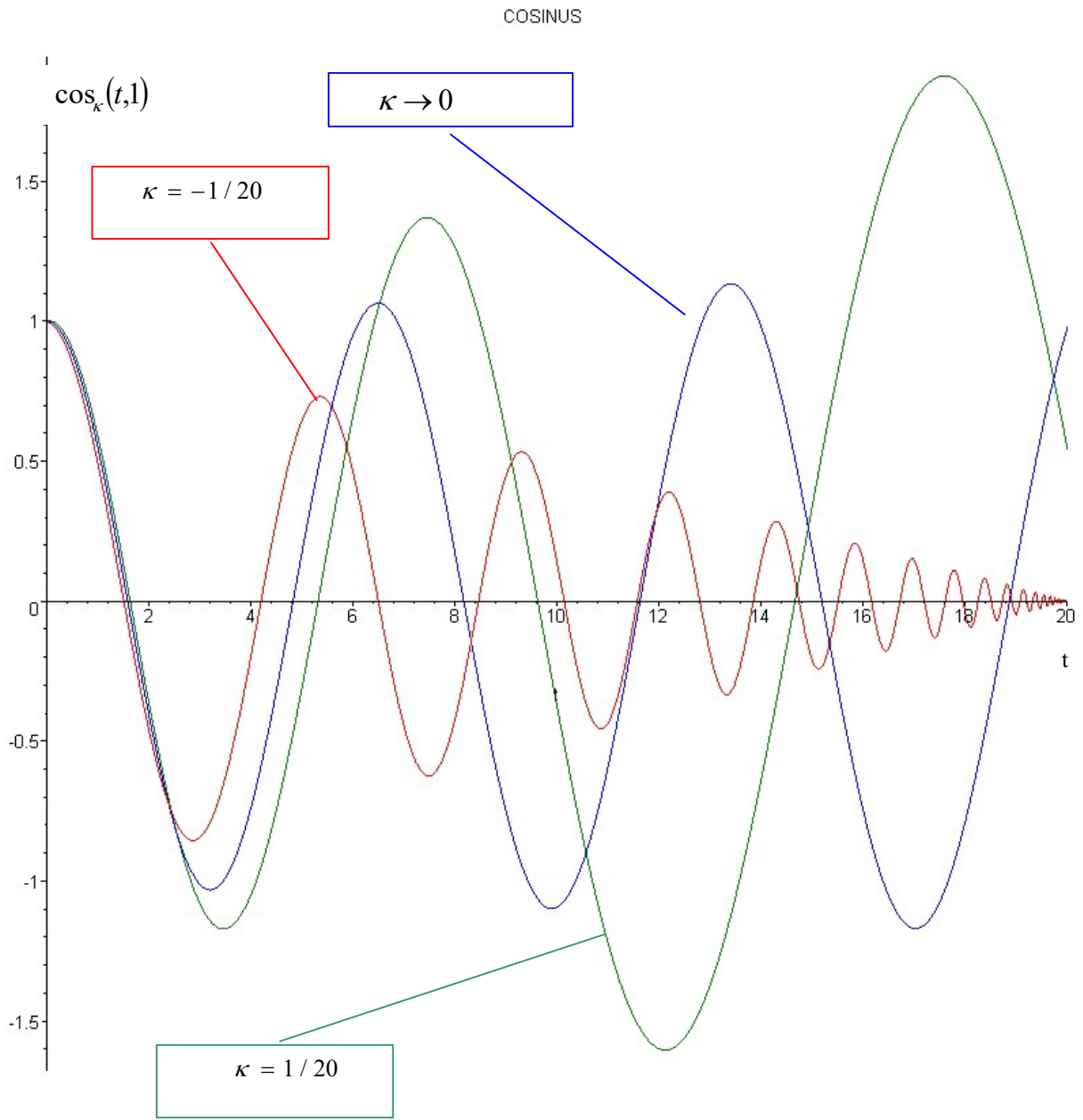

Fig 3. Functions $\cos_\kappa(t,1)$ different values of $\kappa$

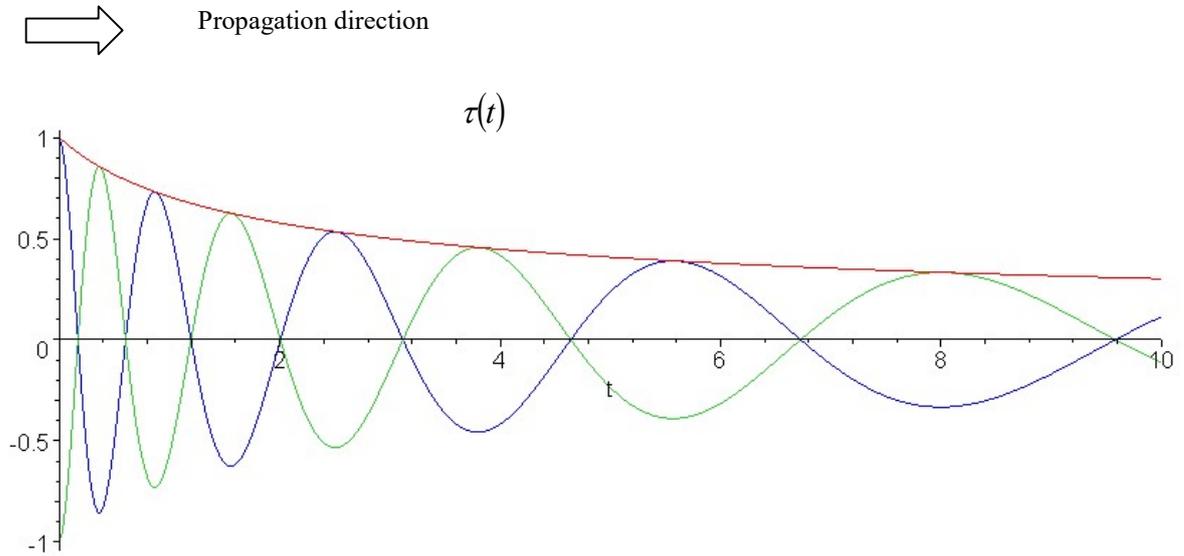

Fig 4 Wave functions $\tau(t)$ for one-dimensional wave equation in the case of "red shift", N=1, $\omega = 1$, $\kappa = 1/100$.

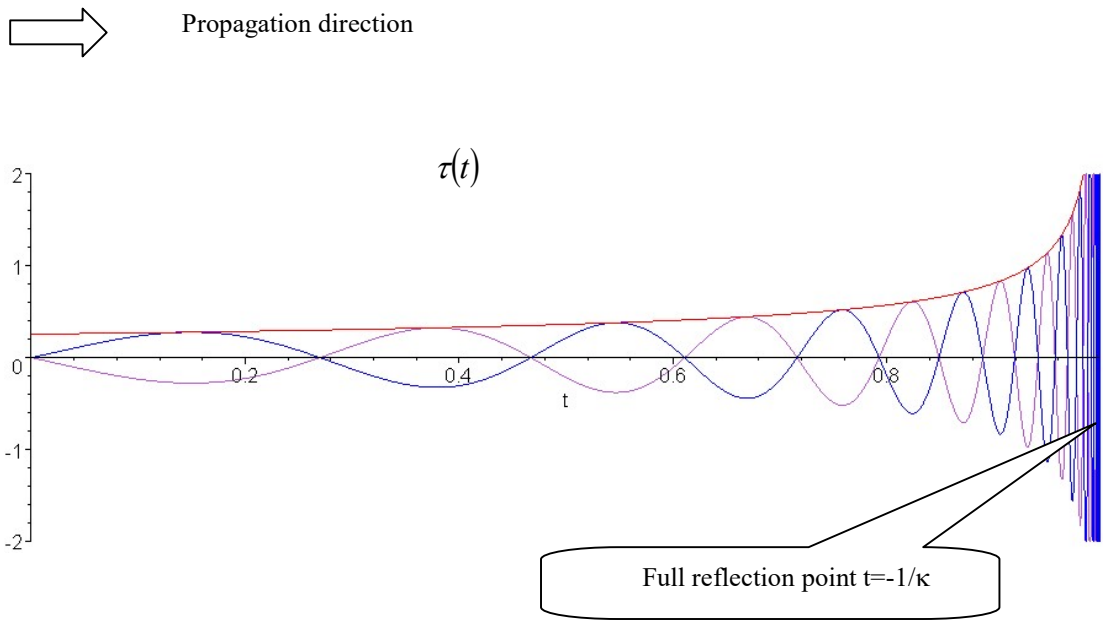

Fig 5. Wave functions $\tau(t)$ for one-dimensional wave equation in the case of "blue shift", N=1, $\omega = 1$, $\kappa = -1/100$.

**List of symbols**

| Symbol | Description |
|---|---|
| ð | operator of generalize derivation ($\kappa$-derivative) |
| $ð/ðt + \tfrac{3}{2}\kappa = Ð$ | Self-adjoint generalized derivation operator |
| $\kappa$ | scaling parameter dimension of an inverse time or inverse length |
| $\exp_\kappa(x,\omega)$ | $\kappa$-exponential function of argument $x$ and eigenvalue $\omega$ |
| $\cos_\kappa(x,\omega), \sin_\kappa(x,\omega)$ | $\kappa$-trigonometric functions of argument $x$ and eigenvalue $\omega$ |
| $\Xi(t,\omega) = \cos(\ln(1+\kappa t)\omega/\kappa)$ | $\kappa$-function $\Xi_\kappa$ of argument $x$ and the eigenvalue $\omega$ (Eq. 16) |
| $\Psi(t,\omega) = \sin(\ln(1+\kappa t)\omega/\kappa)$ | $\kappa$-function $\Psi_\kappa$ of argument $x$ and the eigenvalue $\omega$ (Eq. 16) |
| $\int_{(\kappa)} f(x)dx \stackrel{def}{=} F(x)$ | $\kappa$-indefinite integral ($\kappa$-antiderivative) |
| $F_n(x)$ | indefinite integral ($\kappa$-antiderivative) of the function $f = x^n$ |
| $p^i$ | $\kappa$-canonical momenta |
| $q_i$ | generalized coordinates |
| $L$ | $\kappa$-Lagrange function |
| $L_0$ | ordinary Lagrange function |
| $\Im$ | $\kappa$-Lagrange density |
| $c = 1$ | velocity of propagation of wave motion |
| $N$ | dimension of space where wave propagate ($N = 1,2,3$) |
| $\psi(r,t) = \rho(r)\tau(t)$ | A component of potential |
| $\phi(k)$ | spectral function |
| $\Diamond$ | Non-Leibniz D'Alembert operator |
| $k = \omega/c$ | wave number |